\documentclass[10pt]{iopart}
\usepackage{epsf,float}

\begin{document}

\title{Effect of in-medium properties on heavy-ion collisions}

\author{J\"urgen Schaffner-Bielich\dag\footnote{JSchaffner@bnl.gov}}

\address{\dag\ 
RIKEN BNL Research Center, Brookhaven National Laboratory,
Upton, NY 11973, USA}

\begin{abstract}
The properties of strange hadrons, i.e.\ of kaons and hyperons, in the nuclear
medium are discussed in connection with neutron star properties and
relativistic heavy-ion collisions. 
Firstly, the relevant medium modifications of a kaon in a medium
as provided by heavy-ion collisions is critically examined within a coupled
channel calculation. 
We demonstrate, that particle ratios for kaons are {\em not}
a sensitive probe of in-medium effects while the K$^-$ flow is more suited 
to pin down the K$^-$ optical potential in dense matter. 
Secondly, the interaction between hyperons is studied and may form bound states
which can be produced in relativistic heavy-ion collisions. 
Signals for the detection of strange dibaryons 
by their decay topology and/or in the invariant mass spectra are outlined.
\end{abstract}

\section{Introduction}

We study in the following the in-medium effects of hadrons with
strangeness. The paper is split in two main parts, one dealing with kaons and antikaons, the
other with hyperons. In both parts, we discuss first our basic knowledge of
the underlying kaon/antikaon-nucleon, hyperon-nucleon and hyperon-hyperon
interaction, respectively, and how it relates to an optical potential at normal nuclear
density. Implications for the properties of neutron stars are discussed. 
Then we proceed to apply the possible in-medium effects to heavy-ion
collisions. In the first part about kaons and antikaons, special emphasis will
be put on the fact, that 
the in-medium effects probed in heavy-ion collisions are entirely different from the
ones in the interior of neutron stars. This results in a reinterpretation of
the subthreshold production rates of antikaons as measured at GSI. In the
second part about hyperons, we focus on the possibility of forming bound
states of hyperons in heavy-ion collisions at freeze-out by virtue of a highly attractive
hyperon-hyperon interaction. The properties and decay patterns of these strange exotic
particles are quite unique and make them well distinguishable experimentally from ordinary
hadrons and nuclei.

\section{Kaons in matter}

\subsection{Kaon-nucleon interaction}

The study of kaons in dense matter has attracted much interest in recent
years in particular due to its possible implications for neutron stars. For
the K$^+$ we know from the low-density theorem, that the optical potential is
repulsive in matter 
\begin{equation}
\Pi_{K^+} = t_{K^+N} \cdot \rho_N \quad \longrightarrow \quad
U_{\rm opt}^{K^+} = + 30 \mbox{ MeV } \cdot \rho/\rho_0
\end{equation}
where $t_{K^+N}$ is the (known) elementary K$^+$N scattering amplitude. The
low-density theorem can not be applied to the K$^-$, because of the appearance 
of inelastic channels in dense matter as K$^-$N$\to\Sigma\pi$. The strong
coupling to these inelastic channels, they are about 100 MeV {\em below}
threshold, gives rise to a large imaginary part and to a resonant state just
below threshold so that the isospin zero scattering length is repulsive. 
This resonant state, the $\Lambda(1405)$ resonance, controls the in-medium
properties of the K$^-$. There are two indications, that the optical potential 
for the K$^-$ changes sign in the medium. Firstly, the study of kaonic atoms
reveals that a fit with a nonlinear density dependence can reduce the energy
of the kaon by nearly 200 MeV inside the nucleus \cite{Fried93,Fried94}. Secondly, a
coupled channel calculation for the K$^-$ in the nuclear medium produces also
an attractive potential of $U_{K^-}\approx -100$ MeV at $\rho_0$
\cite{Koch94}, although it is much shallower than the one extracted from
kaonic atoms. Here, the $\Lambda(1405)$ mode is shifted up in the
medium, so that the K$^-$ feels its truly attractive potential. 
The implications for the properties of neutron stars can be quite
drastic. Ignoring the hyperon degrees of freedom, the maximum mass of a
neutron star is considerably reduced \cite{Li97}. Also, the minimum radius
can decrease from values around 12 km to nearly 8 km \cite{GS98L}. These small
radii occur only for relatively large values of the K$^-$ optical potential
which are between the values quoted from kaonic atoms and coupled channel
calculations. In this case, a first order phase transition occurs in the
interior of the neutron star building a mixed phase of an ordinary nucleonic
phase and a kaon condensed phase. The smallness of the radius of a neutron
star might then signal a first order phase transition. But as we will see
below, this is {\em not} a unique feature for kaon condensation inside neutron stars.

Heavy-ion collisions provide an important tool to pin down the possible
in-medium effects for antikaons. Especially, the production rates of K$^-$ at
subthreshold bombarding energies are regarded as a measure of the antikaon
optical potential \cite{Li97}. Now the situation in heavy-ion collisions is completely
different compared to kaonic atoms or neutron stars. In the latter two cases,
the antikaon optical potential is probed at zero momentum and zero
temperature. In heavy-ion collisions, on the contrary, matter is in motion and 
antikaons are produced with a finite momentum relative to the matter rest
frame. In addition, the matter is not only dense but also heated up --- slope
parameters of 90 MeV have been extracted from the hadron spectra. So the
relevant observable in heavy-ion collisions is the optical potential of the
K$^-$ at finite relative momentum and finite temperature.

\subsection{Coupled channel calculations at finite relative momentum and temperature}

We studied the properties of the K$^-$ in such an environment within a coupled 
channel calculation \cite{SKE00}. A simple separable SU(3) symmetric
potential is used whose parameters are fixed by the shape of the
$\Lambda(1405)$ resonance. This also gives a reasonable description of the
various cross sections, although we underestimate the elastic channel and
overestimate the charge exchange reaction K$^-$p$\to$ K$^0$n. In the medium, the
loop integral over the intermediate states ($\bar KN$, $\pi\Sigma$,
$\pi\Lambda$) is modified. In the following, we just discuss the simplest
case, when only the nucleon propagator is changed in the medium. In-medium
effects for the hyperons are found to be negligible
\cite{SKE00}. A selfconsistent treatment of the antikaon, as done in
\cite{Lutz98,Ramos00}, would only strengthen our conclusions about the optical
potential of the antikaon in heavy-ion matter.

\begin{figure}
\begin{center}
\epsfysize=0.30\textheight
\epsfbox{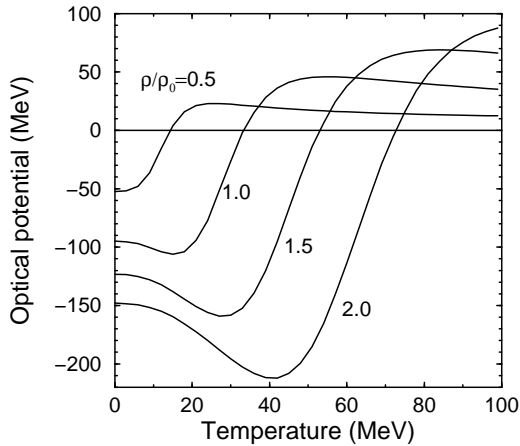}
\end{center}\vspace*{-1cm}
\caption{The optical potential of the K$^-$ as a function of temperature for
  different densities (taken from \cite{SKE00}).}
\label{fig:uopt_temp}
\end{figure}

The temperature dependence of the K$^-$ optical potential is shown in
figure~\ref{fig:uopt_temp} for different densities. One sees, that the K$^-$
optical potential suddenly changes sign for a sufficiently large
temperature. For normal nuclear density $\rho_0$, this happens to be at a
critical temperature of $T_c=32$ MeV. 
For twice normal nuclear density the critical temperature is about twice this
value. In both cases, the critical temperature stays well below the slope
parameter of $T=90$ MeV measured in the relevant heavy-ion experiments at
GSI. Note, that the value of $T_c$ is equal to the nucleon Fermi energy. 
Pauli blocking effects begin to wash out and to get ineffective at $T_c$ for providing an
attractive potential for the antikaon. At large temperatures, the optical
potential saturates at a positive (repulsive) value. Here, the interaction
between K$^-$ and the nucleons can again be described by the low density
theorem as the systems gets diluted by the high temperature.
The repulsive potential at large temperatures 
is then just proportional to the repulsive scattering length times the density,
as can be read off from figure~\ref{fig:uopt_temp}.
We found a similar but less pronounced effect, when incorporating a
finite relative momentum for the K$^-$ optical potential \cite{SKE00}. 
For momenta relevant to subthreshold production of K$^-$ at GSI, the optical
potential turns out to be repulsive, too. We conclude at his point, that the
K$^-$ feels a  much shallower if not repulsive potential in heavy-ion
collisions. Hence, the hitherto used argumentation that a reduced effective
energy of the K$^-$  is responsible for the enhanced subthreshold production
of K$^-$ seen at GSI should at least be reconsidered.

\begin{figure}
\begin{center}
\epsfysize=0.30\textheight
\epsfbox{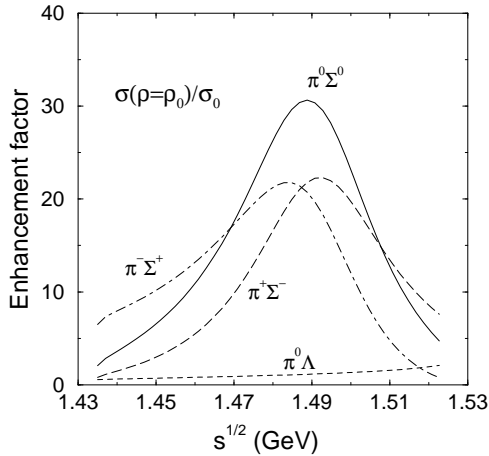}
\end{center}\vspace*{-1cm}
\caption{The enhancement of the production cross section of K$^-$ at normal
  nuclear density compared to the zero density case (taken from \cite{SKE00}).}
\label{fig:sigma_enh}
\end{figure}

There are two comments and a question in order at this point. Firstly, a selfconsistent
treatment of the K$^-$ results also in much shallower optical potentials of
$-30$ to $-40$ MeV at $\rho_0$ even for the zero momentum, zero temperature
case \cite{Lutz98,Ramos00,SKE00}
strengthening our conclusions made above. Recently, a selfconsistent
calculation has been performed by Tolos et al within the Bonn model, which
takes into account all higher partial waves \cite{Tolos2000}. They find a
value of $-80$ MeV for the antikaon optical potential at $\rho_0$, which is still 
lower than the $U_{K^-}= -100$ MeV of the nonrelativistic coupled
channel calculation used above. 
Secondly, the results for finite
momenta have to be taken with care, as p-wave contributions might become
important, which have not been included here (see \cite{Krivo_S2000}). Again
the momentum dependence is much weaker in the Bonn model calculation
including all partial waves \cite{Tolos2000} than in our calculations which takes
into account only the s-wave contributions. Note, that the large temperature
effects seen in figure~\ref{fig:uopt_temp} are unaffected by these comments.

So if the antikaon feels that weak attraction (if at all) in heavy-ion
collisions, how can one explain the enhancement of K$^-$ production rates as
measured at GSI? 

One possible explanation might be provided by studying the in-medium cross
sections for the antikaons.
Indeed, the fact that the $\Lambda(1405)$ mode is shifted up in the medium
directly relates to an tremendously enhanced imaginary scattering amplitude,
i.e.\ an enhanced cross section, at and above the threshold energy. 
We define an enhancement factor of the cross section by the ratio of the cross 
section at normal nuclear density over the one at zero density. The
enhancement factor for the production cross section of the K$^-$ via the
hyperon-pion channels is depicted in figure~\ref{fig:sigma_enh} versus the
colliding energy of the K$^-$p pair. The production cross section
$\Sigma\pi\to K^-p$ is enhanced by an order of magnitude, while the
$\pi^0\Lambda$ channel is barely affected. The reason is, that only the
isospin zero amplitude, where the $\Lambda(1405)$ is sitting, is enhanced in
the medium. The $\pi^0\Lambda$ does not couple to isospin zero, while the
$\Sigma\pi$ does. This explains the difference in the enhancement
factor seen in figure~\ref{fig:sigma_enh}. The dominant peak structure around
$\sqrt{s}=1.49$ GeV stems from the $\Lambda(1405)$ which is shifted well above
threshold in the dense nuclear medium. The position of the peak depends on the 
density. It is noteworthy, that especially the
$\pi^-\Sigma^+$ production channel is enhanced at threshold. Indications of
this medium enhancement can be deduced from K$^-$ absorption on nuclei
\cite{Ohnishi97}. As shown in \cite{SKE00}, finite momentum, finite
temperature and selfconsistency effects are not as crucial for the cross
sections as for the optical potential, so that an enhancement, although
somehow diminished, prevails. 
Hence, the cross section enhancement seems to be a viable tool
to explain the K$^-$ enhancement. As we will see in the following, this is not 
yet the full story.

\begin{figure}
\begin{center}
\vspace*{-0.3cm}
\epsfysize=0.4\textheight
\epsfbox{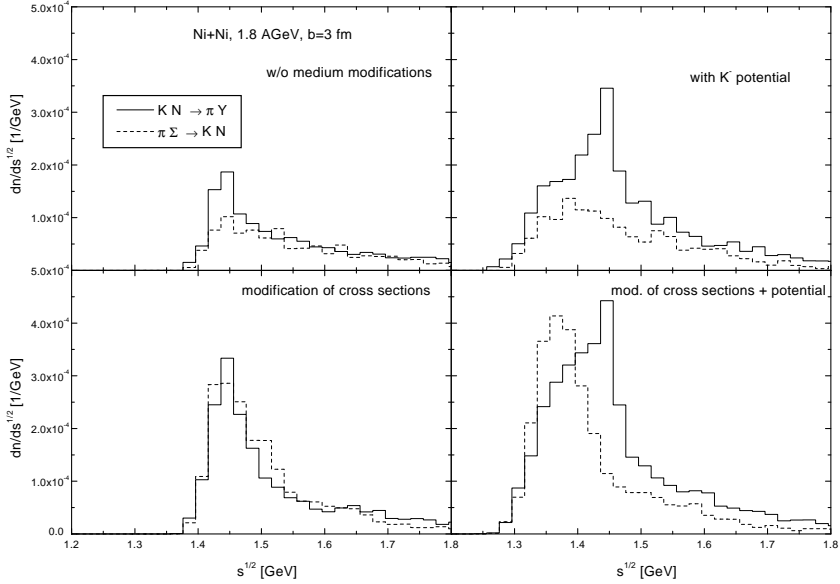}
\end{center}\vspace*{-1cm}
\caption{The collisions spectra for the different versions of the in-medium
  changes for the K$^-$ (taken from \cite{SKE00}). 
  The collision rate increases considerably when
  including effects from the medium modified cross sections.}
\label{fig:coll_spectra}
\end{figure}

\subsection{Confrontation with experimental data on K$^-$ production}

To elucidate the possible influences of in-medium potentials and in-medium
cross sections, a relativistic transport model has to be asked for advice.
Here we use the BUU transport model of \cite{Teis97,Effi99}. Details of the
included strangeness production and absorption channels as well as of the
implemented kaon-nucleon interactions can be found in \cite{SKE00}. 
The model gives a reasonable description of the pion and K$^+$ production
data. 
For the K$^-$, we implement the following three in-medium effects of our
coupled channel calculations in addition to the free case. 
In all these cases, effects from a finite relative momentum and finite
temperature are ignored to study the maximum possible in-medium effect in our approach.
Firstly, an optical potential of the K$^-$
is included which scales 
with density as done in all previous simulations with in-medium effects for
the K$^-$. Note, that this provides the maximal in-medium effect for the K$^-$
optical potential which, as outlined before, is probably weakened at finite
relative momentum. 
Secondly, the in-medium cross section is taken into account.
Thirdly, both effects are added to the simulation. This case gives the maximum
in-medium effect.

Let us begin by looking at the collision spectra in
figure~\ref{fig:coll_spectra}. An in-medium potential for the K$^-$
increases the number of $\bar KN\to Y\pi$ collisions by about a factor two as
more antikaons are produced. The energy shift of the antikaon in the medium
also populates the collisions spectra below the threshold energy. A
modification of the cross section in the medium triples the number of
$\pi\Sigma\to \bar KN$ processes. Finally, the combined effects of in-medium
potential and cross section enhances the collisions spectra by a factor five
below $\sqrt{s}=1.52$ GeV. Note, that a modified cross section enhances both
reactions, the production as well as the absorption reactions, due to detailed 
balance. As the systems is close to equilibrium, 
the enhanced production of antikaons is nearly compensated by the
increased absorption rate.

\begin{figure}
\begin{center}
\vspace*{-0.3cm}
\epsfysize=0.35\textheight
\epsfbox{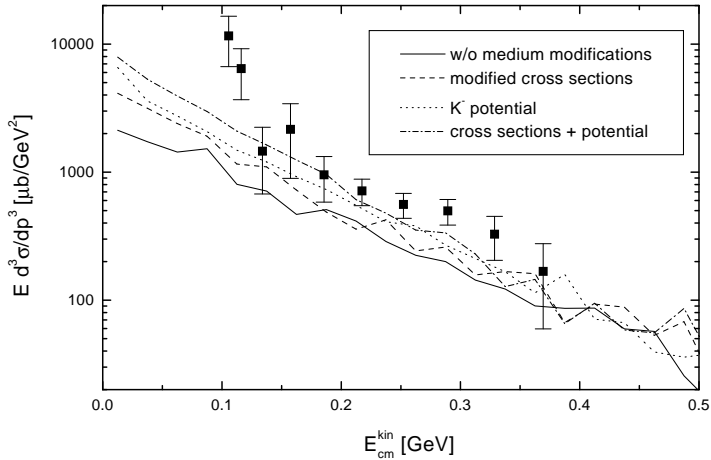}
\end{center}\vspace*{-1cm}
\caption{The transverse spectra of the K$^-$ for Ni+Ni collisions at 1.8 AGeV
  for various choices of medium modifications (taken from \cite{SKE00}). The
  data is from the KaoS collaboration \cite{Kaos97}.}
\label{fig:ekin_spectra}
\end{figure}

This effect shows up also in the final production cross section (see
figure~\ref{fig:ekin_spectra}). The medium effects for the K$^-$ are less drastic
than expected from the sizable in-medium changes of the optical potential
and the cross sections. We see a maximum enhancement of about a factor of
three for low momentum K$^-$ compared to the free case which is much less than
the anticipated order of magnitude enhancement from the in-medium cross
section enhancement or the optical potential. Interestingly, the result without
in-medium effects is already close to the data points of the KaoS
collaboration and the inclusion of in-medium effects just puts the curve
closer to the data.   

\begin{figure}
\begin{center}
\vspace*{-0.3cm}
\epsfysize=0.35\textheight
\epsfbox{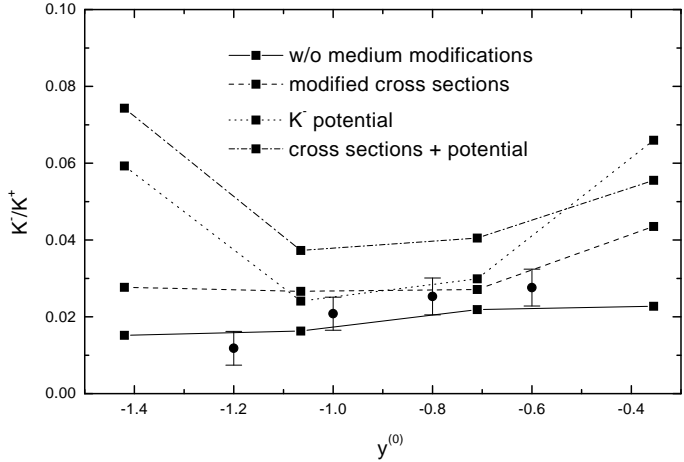}
\end{center}\vspace*{-1cm}
\caption{The ratio of produced K$^-$/K$^+$ for Ru+Ru collisions at 1.69 AGeV
  for $b<4$ fm as a function of beam rapidity $y^{(0)} = y_{\rm beam}/y_{\rm
    CMS} -1$ (taken from \cite{SKE00}) in comparison with the data 
  of the FOPI collaboration \cite{FOPI99}.}
\label{fig:K_ratio}
\end{figure}

The ratio of produced K$^-$ to K$^+$ as measured by the FOPI collaboration is
plotted in figure~\ref{fig:K_ratio}. Again, one notices that the production
rate of K$^-$ is not very sensitive to in-medium effects. Only the combined
maximum effect of a deeply attractive optical potential of the K$^-$ and
an enhanced cross sections results in a significant deviation of the K$^-$/K$^+$
ratio from the one of the free case. While the free case is on top of the data 
points, the in-medium effects seem to overpredict the ratio slightly. 
The in-medium results do not differ significantly from the free case, because 
the system is already so close to chemical equilibrium that
an acceleration of the reaction rates due to larger cross sections or reduced
masses has no further influence on the final production yields.

Our results are in accord with the success of the statistical model 
in the canonical ensemble in describing the kaon and antikaon production data
\cite{Cleymans00,Oeschler_S2000}. 
In this model, no in-medium effect is required to get the
K$^-$/K$^+$ ratio, too. Another hint for the success of the statistical
model might be provided by the collision spectra of figure
\ref{fig:coll_spectra}. The in-medium cross sections enhance the number of
collisions involving the K$^-$ without changing the final number considerably, 
so that a purely statistical analysis becomes applicable. 
Hence, we judge 
from figures~\ref{fig:ekin_spectra} and \ref{fig:K_ratio} that the K$^-$
production numbers are not a very sensitive probe of possible in-medium
effects of the K$^-$ in dense matter.

\begin{figure}
\begin{center}
\vspace*{-0.3cm}
\epsfysize=0.35\textheight
\epsfbox{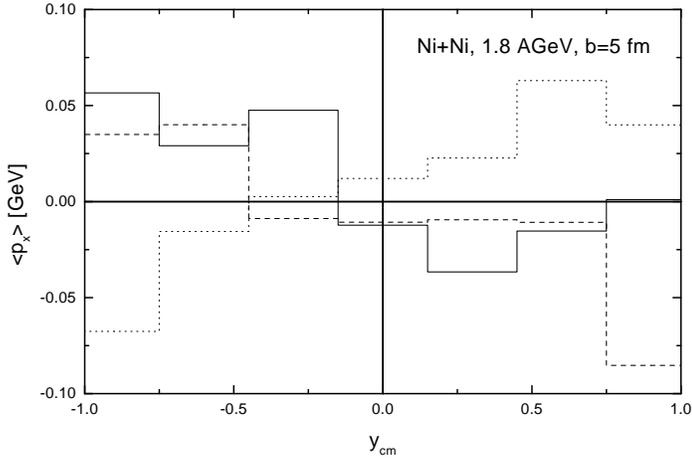}
\end{center}\vspace*{-1cm}
\caption{The flow of K$^-$ in a Ni+Ni collision at 1.8 AGeV and $b=3$ fm with
  and without medium modifications (taken from 
  \cite{SKE00}). The K$^-$'s flow with the nucleons only for an attractive
  optical potential, all other cases studied give an antiflow for the K$^-$.}
\label{fig:K_flow}
\end{figure}

In summary, our investigations seems to indicate that one has to look for more subtle
observables to pin down the possible in-medium effects for the antikaons in a less
ambiguous way. One such observable might be well constituted by the K$^-$
flow pattern. This feature of the K$^-$ flow is demonstrated in
figure~\ref{fig:K_flow}. Without medium modifications and even for a modified
cross section, the K$^-$ shows an antiflow. An attractive K$^-$
optical potential makes the K$^-$ to flow with the nucleons. We found that only
an attractive potential for the K$^-$ switches from an antiflow to a flow of
the K$^-$. The inclusion of temperature effects will also result in an antiflow pattern.
Therefore, a measurement of the flow of the K$^-$ might be better suited to
extract the attractive optical potential of the K$^-$ than the K$^-$
production rates.

\section{Hyperons in matter}

\subsection{Baryon-baryon interactions and strange dibaryons}

In the following we discuss the properties of hyperons in dense matter. 
The detailed study of $\Lambda$-hypernuclei gives important information about
the underlying nucleon-$\Lambda$ interaction. Actually, the $\Lambda$ is the
only hadron, besides the nucleons, where we know for sure the in-medium
properties at normal nuclear matter density as it has been measured to be
$U_\Lambda=-27$ MeV, contrary to e.g.\ the kaons and antikaons. For the other
hyperons the situation is less clear. No bound states of heavy
$\Sigma$-hypernuclei have been found \cite{Bart99} (note that the only known
$\Sigma$-hypernucleus $^4_\Sigma$He is bound by virtue of an attractive
isospin force \cite{Nagae98}). The nucleon-$\Xi$ interaction as extracted
from the few $\Xi$-hypernuclear events is attractive and gives a 
potential depth of $U_\Xi=-20$ to $-25$ MeV at $\rho_0$ \cite{Dover83}. 
Recent measurements of final state interactions of nucleons and $\Xi$
indicate a somewhat smaller value of $U_\Xi=-14$ MeV or less \cite{Khaustov2000}. 
The $\Lambda\Lambda$ interaction is also found to be attractive in the
available double $\Lambda$ hypernuclear data \cite{Dover91}, even more
attractive than the N$\Lambda$ interaction. For the other hyperon-hyperon
interactions ($\Lambda\Sigma$, $\Sigma\Sigma$, $\Lambda\Xi$, $\Sigma\Xi$ and
$\Xi\Xi$) nothing is known at present. Nevertheless, a strongly 
attractive hyperon-hyperon interaction can quite strongly effect the global
properties of a neutron star.  A third solution to the
Tolman-Oppenheimer-Volkov equations can be found for highly attractive
hyperon-hyperon potentials, which has extremely small radii of 6--8 km
\cite{SHSG00}.  Hence, these so called hyperstars have similar small radii as
the ones obtained for kaon condensation in neutron stars. 

Theoretically, the strength of the hyperon-hyperon interaction can not be
deduced from general arguments. Model calculations predict that the
hyperon-hyperon interaction is so attractive in certain channels that it
leads to the existence of bound systems of two hyperons. The most recent
Nijmegen model even suggests that states of $\Sigma\Xi$ and
$\Xi\Xi$ dibaryons might be deeply bound \cite{Stoks99a}.

\subsection{Weak decay patterns and signals in heavy-ion collisions}

Heavy-ion experiments are the only terrestrial source which can
possibly shed light on the issue of the baryon-baryon interactions for hyperons.
Dozens of hyperons can be produced in a single central heavy-ion collisions,
sitting close in phase space. At freeze-out, pairs of hyperons can form and
build a bound strange composite object. This strange dibaryon will then decay
on the timescale of weak interactions. Here, we discuss possible signals, if
such a bound strange dibaryon indeed exists. The weak decay of the strange
dibaryon is modelled by using weak SU(3) symmetry for the parity-violating and 
parity-conserving amplitudes, whose parameters are fixed by the weak decay
amplitudes of the octet hyperons \cite{SMS00}. There are two possible decays
for the dibaryon: mesonic like in the case of free hyperon decay and nonmesonic
without emitting a meson. The latter one is only possible in the medium and
has been studied experimentally by the weak decay of medium to heavy $\Lambda$
hypernuclei. The standard approach to the nonmesonic decay of $\Lambda$
hypernuclei is the meson exchange model \cite{Parreno97}. We extended the
meson exchange model for the weak nonmesonic decay 
to the other hyperons taking into account only the intermediate pion and kaon
exchanges \cite{SMS00}. 
The resulting decay lengths of different strange dibaryons from these model
calculations are in the range of $c\tau=1$--5 cm, which are only slightly shorter than
the ones for the free hyperons.

\begin{figure}
\begin{center}
\epsfysize=0.30\textheight
\epsfbox{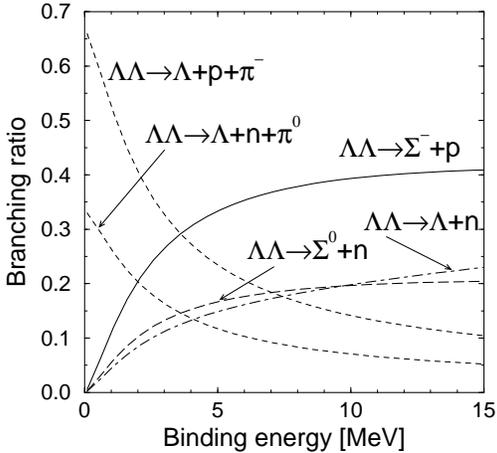}
\end{center}
\vspace*{-1cm}
\caption{Branching ratios for the weak decay of a bound ($\Lambda\Lambda$)
  dibaryon as a function of binding energy (taken from \cite{SMS00}).} 
\label{fig:lala}
\end{figure}

For the two-$\Lambda$ dibaryon, the branching ratio for the various decay channels
are summarized in figure~\ref{fig:lala}. The dominant decay channel for small
binding energies is the free (mesonic) one, as the two $\Lambda$'s are well
separated. For larger binding energies, the dominant decay channel is
$\Lambda\Lambda\to\Sigma^-+{\rm p}$, the same as for the H dibaryon \cite{Don86} one 
is looking for in experiment E896 \cite{Caines99}. 

\begin{figure}
\begin{center}
\epsfysize=0.30\textheight
\epsfbox{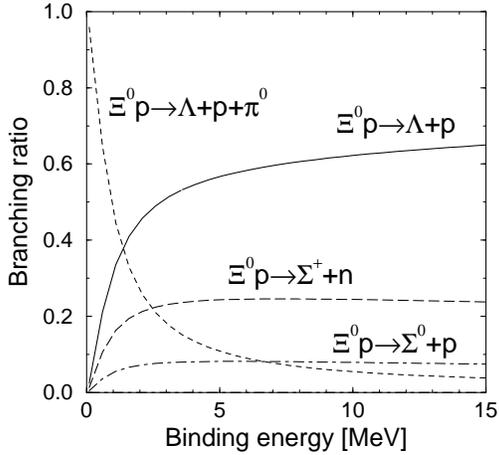}
\end{center}\vspace*{-1cm}
\caption{Branching ratios for the weak decay of a bound ($\Xi^0$p)
  dibaryon as a function of binding energy (taken from \cite{SMS00}).}  
\label{fig:xi0p}
\end{figure}

Figure~\ref{fig:xi0p} shows the case for a bound $\Xi^0$p dibaryon. The
$\Xi^0$p decays mainly into a $\Lambda$ and a proton for binding energies
of only 1.5 MeV and beyond. This decay can in principle be seen either in the
invariant mass spectrum of $\Lambda$'s and protons or directly by the decay
topology. The decay of $\Xi^0$p dibaryon looks like the one for a normal
$\Xi^-$ or $\Omega^-$ hyperon, but instead of a meson a proton is leaving the
first weak vertex. 

Strange dibaryons can have quite exotic properties, as they can have negative
charge while carrying a positive baryon number. There are three possible strange
dibaryons which even have charge $-2$: the $\Sigma^-\Sigma^-$, $\Sigma^-\Xi^-$ 
and $\Xi^-\Xi^-$. Interestingly, all of these have been predicted to be bound
in the recent version of the Nijmegen model \cite{Stoks99a}. The first
candidate has only one possible decay which is mesonic:
$\Sigma^-\Sigma^-\to\Sigma^-+{\rm n}+\pi^-$. As the matrix element of the
mesonic decay is practically unchanged in the medium, we immediately know that 
the decay length of this object is
$c\tau(\Sigma^-\Sigma^-)=c\tau(\Sigma^-)/2=2.2$ cm. The other two candidates
have mesonic as well as nonmesonic decay channels. But in all weak decays,
there is a unique prong as a track with charge $-2$ suddenly splits up into two
tracks with charge $-1$. Such a weak decay is not possible for ordinary
matter. It is in principle also sufficient to measure just the mass of an
object with charge $-2$ as the strange dibaryons are lighter than an
anti-helium nucleus. 

The production rates of strange dibaryons have been estimated for Au+Au collisions
at BNL's Relativistic Heavy Ion Collider using RQMD2.4 with a wavefunction
coalescence \cite{SMS00}. The rates vary between $10^{-2}$ to $10^{-4}$ per
event depending on the strangeness content of the dibaryon. The rapidity
distribution is rather flat, so that dibaryons are even produced at forward
and backward rapidities. Hence, the production rates are reasonably high to search for 
these exotic states in heavy-ion collisions and open the door to the hitherto
unexplored hyperon world. 

\ack
JSB thanks the organizers of this conference for their kind support and 
RIKEN, Brookhaven National Laboratory and the U.S. Department of Energy
for providing the facilities essential for the completion of this work.

\end{document}